\documentclass{article}

\usepackage{microtype}
\usepackage{graphicx}
\usepackage{subcaption}
\usepackage{booktabs}
\usepackage{diagbox}

\usepackage{hyperref}
\usepackage[dvipsnames]{xcolor}

\newcommand{\theHalgorithm}{\arabic{algorithm}}

\usepackage{algorithm}
\usepackage[noend]{algpseudocode}
\newcommand{\mlsysSkipAlgorithmic}{}
\providecommand{\Description}[1]{}
\usepackage[accepted]{mlsys2025}
\usepackage{suppress_notice}
\usepackage{listings}
\usepackage[normalem]{ulem}

\mlsystitlerunning{\sys: A Low-Storage Overhead Vector Index}

\definecolor{col_overview_a}{RGB}{255,204,204}
\definecolor{col_overview_b}{RGB}{204,102,0}
\definecolor{col_overview_c}{RGB}{102,204,0}

\newcommand*\circlea[1]{\tikz[baseline=(char.base)]{%
		\node[white,shape=circle,fill=col_overview_a,draw,inner sep=1pt] (char) {\color{black}\sffamily #1};}}
\newcommand*\circleb[1]{\tikz[baseline=(char.base)]{%
		\node[white,shape=circle,fill=col_overview_b,draw,inner sep=1pt] (char) {\color{black}\sffamily #1};}}
\newcommand*\circlec[1]{\tikz[baseline=(char.base)]{%
		\node[white,shape=circle,fill=col_overview_c,draw,inner sep=1pt] (char) {\color{black}\sffamily #1};}}

\definecolor{col_overview_w}{RGB}{204,102,204} 

\definecolor{col_overview_d}{RGB}{102,153,255} 

\newcommand{\stitle}[1]{\vspace*{0.4em}\noindent{\bf #1\/}}
\newcommand{\sys}{LEANN\xspace}%

\newcommand{\twolevel}{two-level search\xspace}
\newcommand{\dynbatch}{dynamic batching\xspace}
\newcommand{\graphprune}{graph pruning\xspace}
\newcommand{\highdegreeprune}{high-degree preserving graph pruning\xspace}
\newcommand{\graphrecomp}{graph-based recomputation\xspace}

\newcommand{\Twolevel}{Two-Level Search\xspace}
\newcommand{\Dynbatch}{Dynamic Batching\xspace}

\newcommand{\Highdegreeprune}{High-Degree Preserving Graph Pruning\xspace}

\newcommand{\rerankratio}{re-ranking ratio\xspace}

\newcommand{\storageeff}{storage-efficient\xspace}

\newcommand{\shardedmerge}{sharded merging pipeline\xspace}

\newcommand{\personaldevices}{personal devices\xspace}

\usepackage{xspace}
\usepackage{pifont}
\usepackage{enumitem}
\usepackage{amsmath}

\usepackage[capitalise,noabbrev]{cleveref}

\crefformat{section}{#2\S#1#3}
\crefname{appendix}{Appendix}{Appendices}
\Crefname{appendix}{Appendix}{Appendices}
\crefformat{appendix}{#2Appendix~#1#3}
\Crefformat{appendix}{#2Appendix~#1#3}
\crefrangeformat{appendix}{#3Appendices~#1#4--#5#2#6}
\Crefrangeformat{appendix}{#3Appendices~#1#4--#5#2#6}
\crefname{figure}{Figure}{Figures}
\Crefname{figure}{Figure}{Figures}
\crefformat{figure}{#2Figure~#1#3}
\Crefformat{figure}{#2Figure~#1#3}
\crefname{table}{Table}{Tables}
\Crefname{table}{Table}{Tables}
\crefformat{table}{#2Table~#1#3}
\Crefformat{table}{#2Table~#1#3}
\crefname{ALC@line}{line}{lines}
\Crefname{ALC@line}{Line}{Lines}
\crefformat{ALC@line}{#2line~#1#3}
\Crefformat{ALC@line}{#2Line~#1#3}
\usepackage{graphicx}
\usepackage{filecontents}
\usepackage{amsfonts}
\usepackage{xurl}
\usepackage[htt]{hyphenat}
\usepackage{pythonhighlight}
\usepackage{algorithm}
\usepackage[noend]{algpseudocode}
\usepackage{hyperref}
\usepackage{listings}
\usepackage{colortbl}
\usepackage[utf8]{inputenc}
\usepackage{outlines}
\usepackage{booktabs} 
\usepackage{array}
\usepackage{tikz}
\usepackage{subcaption}
\usepackage{pgfplots}
\usepackage{multirow} 
\newcommand{\squishlist}{
 \begin{list}{$\bullet$}
  { \setlength{\itemsep}{1pt}
   \setlength{\parsep}{1pt}
   \setlength{\topsep}{2.5pt}
   \setlength{\partopsep}{0.5pt}
   \setlength{\leftmargin}{1em}
   \setlength{\labelwidth}{1em}
   \setlength{\labelsep}{0.6em}
  }
 }
 \newcommand{\squishend}{
 \end{list}
}

\makeatletter
\renewcommand{\theHALG@line}{\theHalgorithm.\arabic{ALG@line}}
\makeatother

\usepackage{xspace}    
\usepackage[labelfont=bf]{caption}    

\begin{document}

\twocolumn[
\mlsystitle{\sys: A Low-Storage Overhead Vector Index}

\begin{mlsysauthorlist}
  \mlsysauthor{Yichuan Wang$^{\dag}$}{UCB}
  \mlsysauthor{Zhifei Li}{UCB}
  \mlsysauthor{Shu Liu}{UCB}
  \mlsysauthor{Yongji Wu$^{\dag}$}{UCB}
  \mlsysauthor{Ziming Mao}{UCB}
  \mlsysauthor{Yilong Zhao}{UCB}
  \mlsysauthor{Xiao Yan}{CUHK}
  \mlsysauthor{Zhiying Xu$^{*}$}{AWS}
  \mlsysauthor{Yang Zhou}{UCB,UCD}
  \mlsysauthor{Ion Stoica}{UCB}
  \mlsysauthor{Sewon Min}{UCB}
  \mlsysauthor{Matei Zaharia}{UCB}
  \mlsysauthor{Joseph E. Gonzalez}{UCB}
\end{mlsysauthorlist}

\mlsysaffiliation{UCB}{UC Berkeley}
\mlsysaffiliation{CUHK}{CUHK}
\mlsysaffiliation{AWS}{Amazon Web Services}
\mlsysaffiliation{UCD}{UC Davis}

\mlsyscorrespondingauthor{Yichuan Wang}{yichuan\_wang@berkeley.edu}
\mlsyscorrespondingauthor{Yongji Wu}{wuyongji317@gmail.com}

\mlsyskeywords{}

\vskip 0.3in

\begin{abstract}
Embedding-based vector search underpins many important applications, such as recommendation and retrieval-augmented generation (RAG). It relies on vector indices to enable efficient search. However, these indices require storing high-dimensional embeddings and large index metadata, whose total size can be several times larger than the original data (e.g., text chunks). Such high storage overhead makes it difficult, or even impractical, to deploy vector search on personal devices or large-scale datasets.
To tackle this problem, we propose \sys, a storage-efficient index for vector search that recomputes embeddings on the fly instead of storing them, and compresses state-of-the-art proximity graph indices while preserving search accuracy. \sys delivers high-quality vector search while using only a fraction of the storage (e.g., 5\% of the original data) and supporting storage-efficient index construction and updates. On real-world benchmarks, \sys reduces index size by up to 50$\times$ compared with conventional indices, while maintaining SOTA accuracy and comparable latency for RAG applications.
\end{abstract}

]

\printAffiliationsAndNotice{Under Review. *This work does not relate to the position at Amazon. \dag\ Corresponding authors. Email: \texttt{yichuan\_wang@berkeley.edu}, \texttt{wuyongji317@gmail.com}.}

\pagestyle{plain}

\section{Introduction}
Advances in foundation models have led to increasingly powerful embedding models, and \textit{embedding-based vector search} has become a core functionality underpinning many important applications, such as content search~\cite{work-in-progress,yin2024devicers}, personal assistants~\cite{he2019streaming,cai2024recall}, and question answering~\cite{yang2018hotpotqa,joshi2017triviaqa}.
In particular, data objects with complex semantics (e.g., texts, images, videos) are mapped to high-dimensional vectors with an embedding model, so that semantically similar or related objects have a small distance between their embeddings. To retrieve objects from a database, the query object (e.g., a text description) is first embedded as a query vector and then used to search for the top-$k$ most similar vectors. Since exact vector search requires a linear scan in high-dimensional space, \textit{approximate nearest neighbor search} (ANNS) is commonly adopted~\cite{aumuller2020ann}, which returns most rather than all of the top-$k$ neighbors. The result quality of ANNS is typically measured by \textit{recall}, defined as the fraction of ground-truth top-$k$ neighbors that appear in the $k$ retrieved vectors.

\Cref{tab:rag_breakdown_ordered} shows that when retrieval-augmented generation (RAG) is applied to a question answering (QA) dataset, vector search methods such as \textit{Hierarchical Navigable Small World} (HNSW)~\cite{hnsw} yield substantially higher downstream accuracy than traditional keyword-based search approaches like BM25~\cite{msjimmy_bm25}. This is because vector search better retrieves passages that are semantically related to the query intent.

\begin{table}[!t]
\centering
\caption{Storage overhead and runtime statistics of different indexing methods for RAG, evaluated on a \textbf{76 GB} text datastore~\cite{together2023redpajama} and a QA dataset~\cite{kwiatkowski-etal-2019-natural} using the \textit{Qwen3-4B} model on an RTX 4090.}
\label{tab:rag_breakdown_ordered}
\vspace{0.3em}

\resizebox{\linewidth}{!}{%
\begin{tabular}{@{}l|rrr>{\columncolor{gray!10}}r@{}}
\toprule
\textbf{Metrics} & \textbf{BM25} & \textbf{HNSW} & \textbf{PQ} & \textbf{\sys} \\
\midrule
Downstream accuracy (\%) & \textit{18.3} & \textbf{\textit{25.5}} & \textit{17.9} & \textbf{\textit{25.5}} \\
Storage size (GB) & \textit{59} & \textit{188} & \textit{20} & \textbf{\textit{4}} \\
\quad \textit{Index metadata} & \textit{-} & \textit{15} & \textit{15} & \textbf{\textit{2}} \\
\quad \textit{Vectors} & \textit{-} & \textit{173} & \textit{5} & \textbf{\textit{2}} \\
End-to-end latency (s) & \textit{21.36} & \textit{20.95} & \textit{25.45} & \textbf{\textit{23.34}} \\
\quad \textit{Search} & \textit{0.03} & \textit{0.05} & \textit{4.53} & \textit{2.48} \\
\quad \textit{Response generation} & \textit{21.33} & \textit{20.90} & \textit{20.92} & \textit{20.86} \\
\bottomrule
\end{tabular}}%
\vspace{-0.5em}
\end{table}

\stitle{Deploying ANNS: Challenges and Opportunities.} Vector search demands substantial storage, as high-dimensional embeddings and index metadata can be several times larger than the original data~\cite{shao2024scaling}.
\cref{tab:rag_breakdown_ordered} shows that a 76 GB text corpus requires 173 GB for embeddings and 15 GB for the HNSW index, a state-of-the-art graph-based ANN method, thereby more than doubling the data size.
This imposes a high storage burden for many use cases, including RAG workloads on \personaldevices\ and semantic search over large datasets (e.g., logs or documents). In particular, running vector search locally (e.g., on laptops or workstations) is attractive because it preserves privacy and enables offline access without uploading data to the cloud~\cite{wang2024mememo}. However, the storage capacity of \personaldevices\ is often insufficient for large-scale embeddings and indices.

To reduce storage overhead, a common approach is to compress embeddings using lossy vector quantization methods such as product quantization (PQ)~\cite{pq}. Approximate distances can then be computed between the query and the compressed vectors. However, achieving small vector sizes requires a high compression ratio. For example, PQ needs about 35$\times$ compression to reduce the vectors to 5 GB in Table\ref{tab:rag_breakdown_ordered}. At such a high ratio, large quantization errors degrade the downstream accuracy of vector search to levels even below keyword search with BM25. Moreover, the 15 GB HNSW index cannot be compressed using vector quantization and still burdens \personaldevices.

An important observation from \cref{tab:rag_breakdown_ordered} is that in RAG workloads, LLM generation dominates end-to-end latency (i.e., response generation takes over 20s on an RTX 4090, while vector search completes in milliseconds).
This long generation time, common in complex reasoning or agentic tasks, relaxes the strict requirement for search latency.
Since overall latency is bounded by generation, we can trade a small amount of search latency for substantial storage savings, enabling much more compact vector indexes. This is an attractive trade-off for \personaldevices or resource-constrained deployments.
Motivated by this observation, we ask:

\vspace{-0.6em}
\begin{quote}
\textit{Can we design a vector index that dramatically reduces storage overhead while maintaining search accuracy and meeting reasonably relaxed latency requirements?
}
\end{quote}
\vspace{-0.6em}

\stitle{Our solution \sys.} We present \sys as a vector index tailored for storage-constrained environments with both system and algorithmic optimizations.
\sys can reduce the index footprint to below 5\% of the original data while preserving high result accuracy and reasonable retrieval latency.
At its core, \sys is guided by two insights:

The first insight is that state-of-the-art proximity graph indexes (e.g., HNSW, which we build upon) require each query to visit only a small subset of embeddings to find its nearest neighbors. Thus, instead of storing all embedding vectors, \sys recomputes them at query time using the same encoder as in index building. 
However, naive embedding recomputation can lead to significant latency overhead.
To mitigate this, \sys introduces a \twolevel algorithm that uses the inaccurate approximate distances at high compression ratios to prune embedding recomputations. Moreover, \sys also employs a \dynbatch mechanism that aggregates embedding computations across search hops over the proximity graph index to improve GPU utilization and reduce recomputation latency.

While embedding recomputation allows removing  exact vectors, the proximity graph index metadata can still be large, as shown in \cref{tab:rag_breakdown_ordered}. For example, if a node (i.e., vector) has 64 neighbors, each adjacency list takes 256 bytes, which is already 25\% in size over the typical 1 KB document chunk for original data~\cite{shao2024scaling}. Our second insight is that the high-degree nodes in a proximity graph are visited much more frequently than the low-degree nodes and thus are more important for vector search. Hence, \sys applies a \highdegreeprune strategy, which removes the low utility edges of low-degree nodes while preserving the edges of high-degree ``hub” nodes~\cite{munyampirwa2024down}. This substantially reduces index size without sacrificing search accuracy and efficiency.

Besides the two key designs, \sys incorporates a storage-efficient \shardedmerge index building strategy, which ensures that storage consumption never exceeds a small budget even when building the index for a large dataset. In addition, \sys also supports updating the compressed index (e.g., adding new data). This significantly reduces the update time while remaining storage-efficient.

We implement \sys\footnote{Code repository: \url{https://github.com/yichuan-w/LEANN}.} on top of FAISS~\cite{faiss}, one of the most popular frameworks for ANNS, and evaluate it across four information retrieval (IR) benchmarks and beyond. Our experiment platforms include a server with NVIDIA RTX 4090 GPU~\cite{nvidia4090} and an M1-based Mac~\cite{mac}. The results show that \sys reduces storage consumption by more than 50$\times$ compared to state-of-the-art vector indexes while maintaining result accuracy. When applied to RAG tasks, \sys incurs about 10\% end-to-end latency overhead (see \cref{tab:rag_breakdown_ordered}).

To summarize, we make the following contributions:  
\squishlist
    \item We are the first to study the storage challenge of vector search and design \sys, a novel \storageeff index that performs on-the-fly embedding recomputation and applies \highdegreeprune to reduce storage while preserving accuracy.
    \item We incorporate a suite of optimizations, including \twolevel, \dynbatch, storage-constrained index building, and efficient index update, to make the entire pipeline both fast and storage efficient.
    \item We demonstrate that \sys achieves over 90\% top-3 recall within one second while using less than 5\% of the raw data storage, maintaining comparable latency for RAG workloads.
\squishend

\section{Background}

\label{sec:bg}

\stitle{Vector search.} To retrieve semantically related or similar objects for unstructured data (e.g., texts, images, videos), vector search is widely used. In particular, given a vector dataset \( \mathcal{X} = \{x_1, x_2, \cdots, x_N\} \subset \mathbb{R}^d \) and a query vector \( q \in \mathbb{R}^d \), vector search finds the top-$k$ vectors in $\mathcal{X}$ that are the most similar to $q$, i.e., 
\begin{small}
\begin{equation}
   |\mathcal{S}_q|=k \  \text{with} \ \Vert q-x_i\Vert\le \Vert q-x_j\Vert \ \forall x_i\in \mathcal{S}_q, x_j\in \mathcal{X}\setminus \mathcal{S}_q.
\end{equation}
\end{small}
The similarity function can also be the inner product or cosine similarity, where larger values indicate higher similarity. However, due to the curse of dimensionality in high-dimensional spaces, exact vector search requires a linear scan~\cite{wang2021comprehensive}, which is costly for large datasets. As such, approximate nearest neighbor search (ANNS) is commonly used~\cite{hnsw,ivf}, which trades minor result inaccuracies for substantially lower query latency. The result quality of ANNS is usually measured by \textit{recall}, which is the fraction of ground-truth top-$k$ neighbors that are contained in the set $\mathcal{S}'_q$ of returned approximate neighbors, i.e., 
\begin{equation}
Recall@K=|\mathcal{S}_q \cap \mathcal{S}'_q|/k.
\end{equation}
Applications such as RAG typically require a high recall (e.g., $\geq 0.9$) for good performance ~\cite{shen2024understandingsystemstradeoffsretrievalaugmented}.

Indexes are essential for the efficiency of vector search by confining the distance computations to a small portion of vectors. The storage cost of a vector index consists of two components, i.e., the vectors and the index metadata. Two types of vector indexes are the most popular, i.e., IVF~\cite{ivf}) and proximity graph~\cite{hnsw,nsg,diskann}. IVF groups the vectors into clusters and represents each cluster with a center vector, and a query first scans the centers and then checks the vectors in a few most similar clusters. Proximity graph connects similar vectors to form a graph and conducts vector search via a best-first traversal on the graph. Although IVF is cheaper to build and requires smaller space to store the index structure, proximity graph achieves the SOTA efficiency for vector search in that it requires much fewer distance computations~\cite{diskann}.

\begin{algorithm}[!t]
	\caption{Best-First Search on Graph-based Index}
	\label{alg:best-search}

	\begin{algorithmic}[1]
		\State \textbf{Input:} Graph $G$, query $q$, entry point $p$, result count $k$, queue size $ef$
		\State \textbf{Output:} $k$ nearest neighbors to $q$
		\State Init size-$ef$ priority queue $C$ with $(p, \operatorname{Dist}(q, x_p))$
		\While{$C$ has unvisited node} \label{alg:best-search:line4}
		\State Read the closest but unvisited node $u$ in $C$\label{line:explore}
        \State Mark $u$ as visited
		\For{each neighbor $v$ of $u$ in $G$}
		\If{$\operatorname{Dist}(q, x_v)$ is not computed}
		\State Extract embedding $x_v$ for $v$\label{line:extract}
		\State Try to insert $(v, \operatorname{Dist}(q, x_v))$ into $C$
		\EndIf
		\EndFor
		\EndWhile
		\State \Return The $k$ nodes with the smallest distances in $C$
	\end{algorithmic}
\end{algorithm}

\stitle{Best-first search on proximity graph.} The variants of proximity graph index (e.g., HNSW~\cite{hnsw}, NSG~\cite{nsg}, Vamana~\cite{diskann}) differ in their edge connection rules, but the query processing algorithm is similar. \Cref{alg:best-search} illustrates the best-first search on the proximity graph.
The search maintains a bounded priority queue~$C$ of candidate nodes,
ordered by their distances to the query~$q$.
At each exploration step (Line~\ref{line:explore}), the algorithm \emph{reads} (but does not remove) the closest unvisited node~$u$ from~$C$
and explores its neighbors.
For each neighbor whose distance has not been computed,
the algorithm extracts the embedding, computes its distance to~$q$,
and inserts the neighbor into~$C$ if the queue is not full or
if the neighbor is closer than the tail entry of~$C$.
The parameter~$ef$ bounds the queue size and acts as a \emph{quality knob}:
a larger~$ef$ improves recall at the cost of more distance computations.
The search terminates once all the nodes in~$C$ have been visited. Empirically, graph-based indexes achieve high recall with only $\mathcal{O}(\log N)$ embedding extractions and distance computations. This is because the graph traversal can quickly converge on the neighbors of the query by moving to more similar neighbors in each step.

\section{\sys Overview}\label{sec:overview}
\begin{figure*}[!t]
	\centering
	\includegraphics[width=0.8\textwidth]{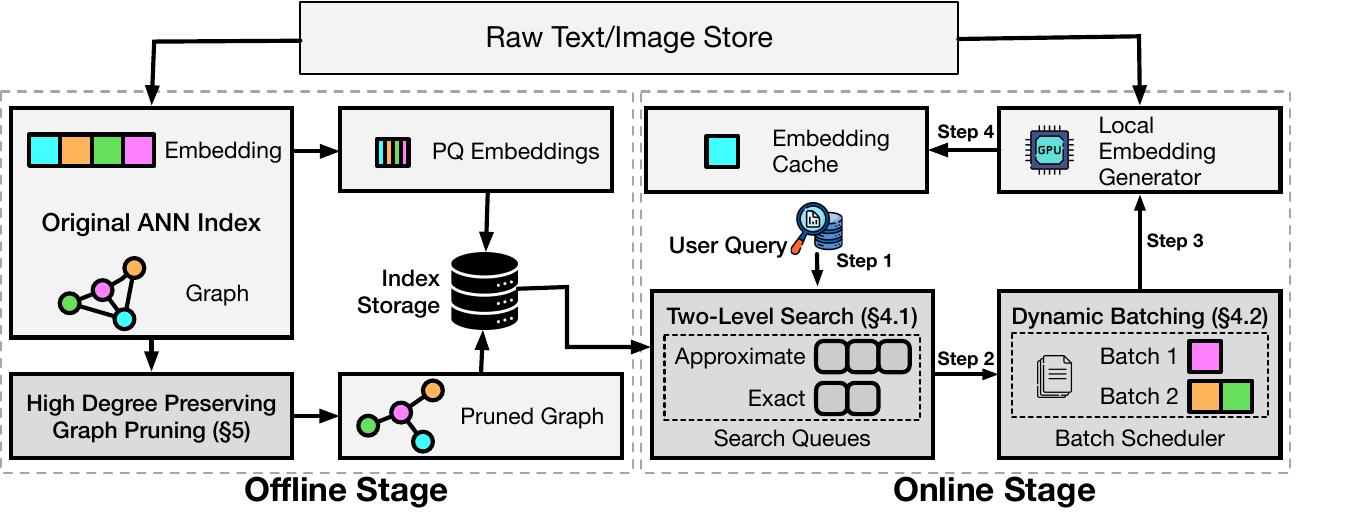}
	\vspace{-1mm}
	\caption{\sys System Diagram. The system combines \highdegreeprune for minimal storage footprint with \graphrecomp and \twolevel with \dynbatch for efficient query processing (Steps 1-4).}
	\label{fig:workflow}
	\vspace{-2mm}
\end{figure*}

\Cref{fig:workflow} shows the end-to-end workflow of \sys, which includes offline index construction and online query serving.

\stitle{Offline stage}
Given a dataset of items, such as chunked unstructured text, \sys computes embeddings and builds a graph-based vector index. 
To minimize storage, it applies a \graphprune algorithm that preserves high-degree nodes~(\cref{sec:pruning}) and discards dense embeddings, retaining only the pruned graph structure. 
During construction, \sys also builds a lightweight product quantization (PQ) table that stores approximate embeddings for fast distance estimation during query processing~(\cref{sec:latency_optimize}). 
Optionally, if a peak storage budget is specified, \sys adopts a graph partitioning-based build strategy~(\cref{sec:build}) to keep the storage footprint within this bound by constructing and merging shards sequentially. 
At the end of the offline stage, \sys persists two compact components: 
(i) the pruned graph adjacency lists and 
(ii) the PQ-compressed embedding table.

\stitle{Online stage.}
When a query arrives, \sys searches over the pruned graph using \cref{alg:best-search}. 
To accelerate query processing, \sys employs a \twolevel strategy that first computes lightweight approximate distances using PQ embeddings and then recomputes exact embeddings on demand for the most promising candidates via the local embedding generator. 
During recomputation, \sys applies \dynbatch to group multiple candidate nodes across exploration steps, improving GPU utilization and reducing end-to-end latency.
Finally, the system ranks all visited nodes by their exact distance to the query and returns the top results to the downstream task. 
\sys also provides a lightweight update pipeline for dynamic index maintenance~(\cref{sec:build}) and, if disk capacity allows, an optional embedding cache to store frequently accessed nodes and avoid redundant recomputation.

\stitle{Storage composition.}
Across both stages, \sys stores compact structures. For $N$ data chunks (nodes), the pruned graph requires $O(N \times |D|)$ integer entries, where $|D|$ denotes the average node degree. The PQ table employs a 100$\times$ smaller codebook than the original FP32 embeddings, occupying $O(4N \times \textit{dim}/100)$ bytes (e.g., $\textit{dim}=768$). Together, these components reduce storage by up to 50$\times$ compared to conventional dense indexes.

\stitle{Use cases.} \sys can conduct vector search on user devices (e.g., laptops and personal servers), on which storage is highly limited. Our experiment evaluation also focuses on this use case. \sys may also be used for data lakes, which contain many datasets, and some cold datasets are queried infrequently~\cite{mageirakos2025cracking}. Storing indexes for these cold datasets incurs high space overheads, while recomputing embeddings for them is inexpensive due to low query frequency. Similarly, \sys can handle datasets whose embeddings have skewed access patterns, e.g., for recommendation and content search, popular entries are more likely to become the results of vector search~\cite{mohoney2023high}. For these datasets, \sys may store exact vectors for the popular entries and use embedding recomputation for the cold entries to reduce storage.   

\section{Graph-based Recomputation} 

\label{sec:latency_optimize}
In this section, we introduce our efficient recomputation pipeline, which reduces the number of nodes involved in recomputation (\cref{sec:doubleq}) and fully utilizes GPU resources during the process (\cref{sec:batch}).

\subsection{\Twolevel with Hybrid Distance}
\label{sec:doubleq}
\begin{algorithm}[!t]
    \caption{Two-Level Search}
    \begin{algorithmic}[1]
        \State \textbf{Input:} query $q$, entry point $p$, re-ranking ratio $\alpha$, result size $k$, search queue length $ef$
        \State \textbf{Output:} $k$ closest neighbors to $q$
        \State Init size-$ef$ priority queue $EQ$ with $(p, \operatorname{Dist}(q, x_p))$
        \State Init empty approximate priority queue $AQ$ 
        \While{$EQ$ has unvisited node}
            \State Read the closest unvisited node $u$ from $EQ$
            \State Mark $u$ as visited
            \For{each neighbor $v$ of $u$}
                \If{approximate distance to $q$ not computed}
                \State Extract approximate embedding $\tilde{x}_v$ for $v$
                \State Insert $(v, \operatorname{Dist}(q, \tilde{x}_v))$ into $AQ$  \label{line:approx_dis}
                \EndIf
            \EndFor
            \State $C \gets$ top $\alpha\%$ candidates in $AQ$, excluding EQ \label{line:extract_top}
            \For{each $c \in C$}
            \State Recompute embedding $x_c$ \label{line:recompute}
                \State Try to insert $(c, \operatorname{Dist}(q,x_c))$ into $EQ$
            \EndFor
        \EndWhile
        \State \Return The $k$ nodes with smallest distances in $EQ$
    \end{algorithmic}
\label{alg:two_level}
\end{algorithm}

\stitle{Motivation.}
\sys stores PQ codes for all vectors to enable approximate distance computation. Existing systems such as DiskANN search the proximity graph using these approximate distances and then re-rank the top candidates with exact distances, e.g., re-ranking the top-100 approximate neighbors for top-10 results. However, this approach is problematic for \sys because our PQ codes use a high compression ratio for compact storage, leading to large quantization errors. In particular, the approximate distances can lead the graph traversal to detours by visiting sub-optimal candidates, which prolongs search time. Moreover, some ground-truth neighbors may be missed due to sub-optimal candidates, and re-ranking more approximate neighbors will not improve recall in this case (see \cref{fig:acc}). To tackle this problem, we interleave approximate and exact distance computations rather than isolating them as in existing systems. Specifically, we use exact distances to select candidates to visit so that the graph traversal maintains high quality, while approximate distances are used to prune unnecessary exact computations, achieving accuracy and efficiency at the same time.

\stitle{Solution.}
\Cref{alg:two_level} outlines the complete procedure. At each exploration step, \sys first computes approximate distances for all neighbors using PQ (Line~\ref{line:approx_dis}) and maintains an approximate queue ($AQ$) that stores these values for all explored nodes.  
Instead of recomputing every neighbor's embedding, we define a \rerankratio~$\alpha$ and extract the top $\alpha\%$ of nodes from $AQ$, excluding those already in the exact queue ($EQ$).  
The selected subset $C$ (Line~\ref{line:extract_top}) is then recomputed exactly, and each node is inserted into $EQ$ for further exploration.\footnote{$AQ$ tracks all previously visited nodes, allowing \sys to revisit earlier neighbors that become more promising as the search progresses.}  
This hybrid strategy significantly reduces recomputation without sacrificing accuracy.

\stitle{Discussion.}
In practice, \sys uses a PQ table with a 100$\times$ smaller codebook, representing embeddings in $O(4N \times \textit{dim}/100)$ bytes (with $\textit{dim}=768$ in our setup).
These approximate distances provide an inexpensive yet effective signal for early filtering, and recomputation is reserved only for a small fraction of top-ranked candidates.  
Although PQ introduces quantization errors, selective exact recomputation restores ranking fidelity and ensures retrieval quality.  
The method generalizes easily to other forms of approximation, such as using distilled embedding models or link-and-code representations~\cite{douze2018link}.

\subsection{\Dynbatch for Recomputation}
\label{sec:batch} 
\stitle{Motivation.}
In the naive approach, embeddings are recomputed one by one for each neighbor node, as shown in Line~\ref{line:recompute} of \cref{alg:two_level}. 
To better utilize the GPU, \sys batches all neighbor nodes within an exploration step so their embeddings are recomputed together. 
However, even with this optimization, each batch remains small---limited by the degree of the current node~$u$. 
The problem becomes more pronounced in the \twolevel algorithm (Line~\ref{line:extract_top}), where the candidate set per step is even smaller.

\stitle{Solution.}
To further improve GPU utilization, \sys introduces a \dynbatch strategy that relaxes the strict data dependency in best-first search (\cref{alg:best-search}). 
While this introduces slight staleness in the exploration order, it enables batching across multiple exploration steps, increasing the effective batch size and improving throughput.

Specifically, \sys dynamically collects a group of the closest candidates from the priority queue. 
The algorithm accumulates nodes requiring recomputation until a target batch size is reached (e.g., 64), which can be efficiently determined through lightweight offline profiling. 
This \dynbatch mechanism integrates naturally with the \twolevel strategy described in \cref{sec:doubleq}: in practice, \sys accumulates nodes in the set $C$ across iterations until the predefined batch threshold is reached, then recomputes embeddings for all nodes in $C$ together.

This \dynbatch approach allows \sys to batch nodes across multiple graph exploration steps regardless of individual node degrees, trading off slight staleness for significantly improved GPU utilization compared to single-step processing.

\section{Compact Graph Structure}
\label{sec:pruning}

\begin{figure}[t]
	\centering

	\begin{subfigure}[b]{\columnwidth}
		\centering
		\includegraphics[width=\linewidth]{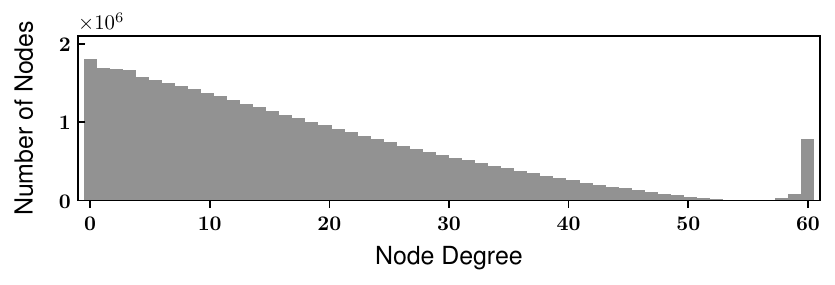}
		\caption{Node (out)-degree distribution}
		\label{fig:node_distribution}
	\end{subfigure}

	\begin{subfigure}[b]{\columnwidth}
		\centering
		\includegraphics[width=\linewidth]{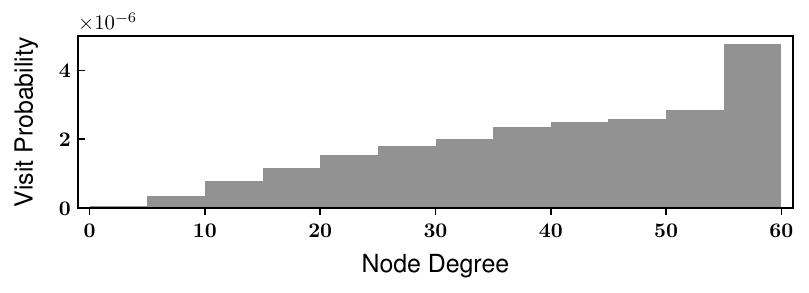}
		\caption{Node access probability.}
		\label{fig:node_access_prob}
	\end{subfigure}


	\caption{HNSW graph analysis reveals skewed access and degree distributions, with node degrees capped at 60 by HNSW.}
	\label{fig:skewnode}
	\Description{}
\end{figure}

With the \twolevel and \dynbatch mechanisms optimizing recomputation latency, we now examine how \sys further reduces storage overhead in graph index metadata through a \highdegreeprune algorithm.
As noted in \cref{sec:overview}, although \sys eliminates the need to store exact embeddings by recomputing them at query time, the graph metadata that guides the search still incurs significant storage cost (see \cref{tab:rag_breakdown_ordered}).
In fact, even with embeddings, the index metadata alone can exceed 30\% of the total storage~\cite{severo2025lossless}.

\stitle{Problem Formulation.}
Given a disk usage constraint \(B\), \sys aims to prune the graph index so that the metadata storage remains within budget while maintaining retrieval accuracy.
Formally, the optimization problem is:

{\small
\vspace{-1.2em}
\begin{equation}\label{equ:nodestorage}
	\begin{alignedat}{2}
		\min \quad
		 & T(G_1)            &  & = \sum_{i=1}^{ef} |V_i|                                \\[3pt]
		\mathrm{s.t.} \quad
		 & \text{Space}(G_1) &  & = \sum_{v \in V(G_1)} \deg(v)\, s_{\text{edge}} \le B, \\[3pt]
		 & \text{Acc}(G_1)   &  & \ge \tau
	\end{alignedat}
\end{equation}
\vspace{-1.2em}
}

Here, \(G_1\) is the pruned graph, and \(|V_i|\) is the number of nodes recomputed in each exploration step during search using \(G_1\).
A smaller \(T(G_1)\) indicates fewer recomputations and thus lower query latency.
\(\text{Space}(G_1)\) denotes the metadata size of the graph, stored in a compressed sparse row (CSR) format, which records each node's outgoing neighbor IDs.
\(\deg(v)\) is the out-degree of node \(v\), and each stored neighbor ID takes 4~bytes.
The goal is to minimize recomputation cost while keeping the graph within the storage budget \(B\) and accuracy (recall) above the threshold \(\tau\).

\stitle{Motivation.} There are two naive ways to shrink the proximity graph index: (1) randomly removing edges, and (2) lowering the degree limit for each node.
Both approaches significantly degrade search accuracy even under mild size reductions, as shown in \cref{fig:graph_prune_ablation}, because they harm graph connectivity, which is crucial for effective traversal. From \cref{fig:skewnode}, we observe that the edges are not equally important:  a small fraction of nodes have high degrees (i.e., approaching or at the degree limit), and these nodes are accessed much more frequently than the low-degree nodes. These high-degree nodes essentially serve as the ``navigation hubs'' for graph traversal, and similar phenomena are also observed in~\cite{munyampirwa2024down}. As such, we preserve the edges for the high-degree nodes to ensure good navigability of the proximity graph and conduct edge pruning for the low-degree nodes.

\begin{algorithm}[t]
	\caption{\Highdegreeprune}
	\begin{algorithmic}[1]
		\State \textbf{Input:} Original graph $G$ with vertex set $V$; construction queue length $efC$; maximum degree $M$ for high-degree nodes; lower degree $m$ for others ($m < M$); proportion of high-degree nodes $\beta$
		\State \textbf{Output:} Pruned graph $G_1$
		\State Init $D[v] \gets \deg(v)$ for all $v \in V$; $G_1 \gets \emptyset$
		\State $V^* \gets$ nodes with the top $\beta\%$ highest degrees in $D$ \label{line:select_important}
		\For{$v \in V(G)$} \Comment{Construct $G_1$}
		\State $W \gets$ Search($v$, $efC$) \Comment{See \cref{alg:best-search}}
		\If{$v \in V^*$} $M_0 \leftarrow M$ \label{line:original}
		\Else{} $M_0 \leftarrow m$ \label{line:low}
		\EndIf
		\State Select up to $M_0$ neighbors from $W$
		\State Add bidirectional edges between $v$ and neighbors
		\State If $\deg(u)>M$ for any neighbor $u$, shrink to $M$ \label{line:preserve}
		\EndFor
	\end{algorithmic}
	\label{alg:construct}
\end{algorithm}

\stitle{Solution:}
Our key insight is that preserving a small set of hub nodes is sufficient to maintain search performance.
Following prior work~\cite{hmann,munyampirwa2024down}, high-degree nodes serve as the backbone of the graph’s connectivity; thus, \sys focuses on retaining these hubs while reducing the overall number of edges.
\Cref{alg:construct} outlines this \highdegreeprune strategy.

We assign degree thresholds based on node importance: most nodes are limited to a lower degree \(m\), while a small fraction of nodes (\(\beta\%\)) can retain up to \(M\) connections (Line~\ref{line:original}). 
Empirically, we set \(m = M / 5\) and determine \(M\) for a given storage budget \(B\) through offline profiling. 
We use node degree as a proxy for node importance and select the top \(\beta\%\) of nodes by degree (Line~\ref{line:select_important}).
Preserving only the top 2\% of high-degree nodes significantly reduces edge count while maintaining high retrieval accuracy.

Moreover, while we restrict the number of outgoing connections when a node is first inserted into the graph (Line~\ref{line:low}), we allow all nodes to form bidirectional links with newly inserted nodes up to the higher threshold \(M\) (Line~\ref{line:preserve}), instead of the lower limit \(m\).
This design ensures that each node retains the opportunity to connect with high-degree hub nodes, thereby preserving graph navigability with minimal impact on search quality.

\section{Index Building and Update} \label{sec:build}
\stitle{Storage-Efficient Index Build.}
The naive index construction in \sys requires precomputing embeddings for all objects to build the graph structure. 
Although these embeddings are discarded afterward to reduce storage at query time, the peak storage usage during construction can still be substantial. 
To address this, \sys introduces a simple yet effective \shardedmerge strategy that builds the index efficiently under a user-specified storage constraint while preserving graph quality. 
The \shardedmerge process consists of three stages:
\circlec{1}~\textit{Soft assignment with k-means.}
We first run \textit{k-means} on a small subset of the corpus to obtain \(k\) centroids. 
Each object is then embedded and assigned to its two nearest centroids. 
This is performed sequentially; after assignment, embeddings are immediately discarded, and only the two-centroid mapping for each passage is retained.
\circlea{2}~\textit{Shard-wise graph construction.}
After the assignment, we build the graph index separately for each of the \(k\) shards. 
For each shard, embeddings are recomputed, the graph is constructed, and the embeddings are discarded. 
Since each passage belongs to two shards, the merged graph achieves good global connectivity.
\circleb{3}~\textit{Graph merging.}
We then merge the \(k\) shard graphs into a single structure. 
For nodes appearing in two shards, we assign the higher of their two HNSW levels as the final layer. 
For lower layers, we merge their edge lists and randomly drop edges when the node degree exceeds \(M\). 
This heuristic yields a well-connected, high-quality graph (see \cref{fig:effi-build} in \cref{app:more-ablation}); more advanced merging strategies, such as RNG-based pruning~\cite{rng}, are left for future work.

\stitle{Efficient Index Update.} We enable efficient updates in \sys through a series of optimizations that substantially reduce computational and storage overhead.
For a single update request, the naive recomputation procedure has a complexity of $\mathcal{O}(M \cdot efC + efC^2 + M^3)$, 
arising from repeated embedding calculations and neighbor maintenance. 
These three terms correspond to neighbor search, neighbor selection, 
and reverse-edge selection and updates. 
\sys improves efficiency through lightweight embedding, caching and a simplified selection strategy, 
eliminating redundant computations and reducing the total cost to $\mathcal{O}(M \cdot efC)$ 
while preserving graph connectivity and quality. 
For deletions, \sys employs soft deletion by marking nodes as inactive rather than removing them from the graph structure, preserving connectivity while avoiding costly graph reorganization.
Details are provided in \cref{sec:sys-add}.

Beyond single-node insertion, \sys supports \textit{batched add} operations with system-level optimizations. 
Incoming embeddings are temporarily buffered, and upon receiving a query, \sys scans the buffer and merges its results with those from the existing graph. 
The buffered entries are then inserted asynchronously, amortizing update costs. 
This design minimizes computation and peak storage usage while maintaining low search latency.

\section{Evaluation}
\label{s:eval}
\begin{figure}[!t]
	\centering
	\includegraphics[width=0.48\textwidth]{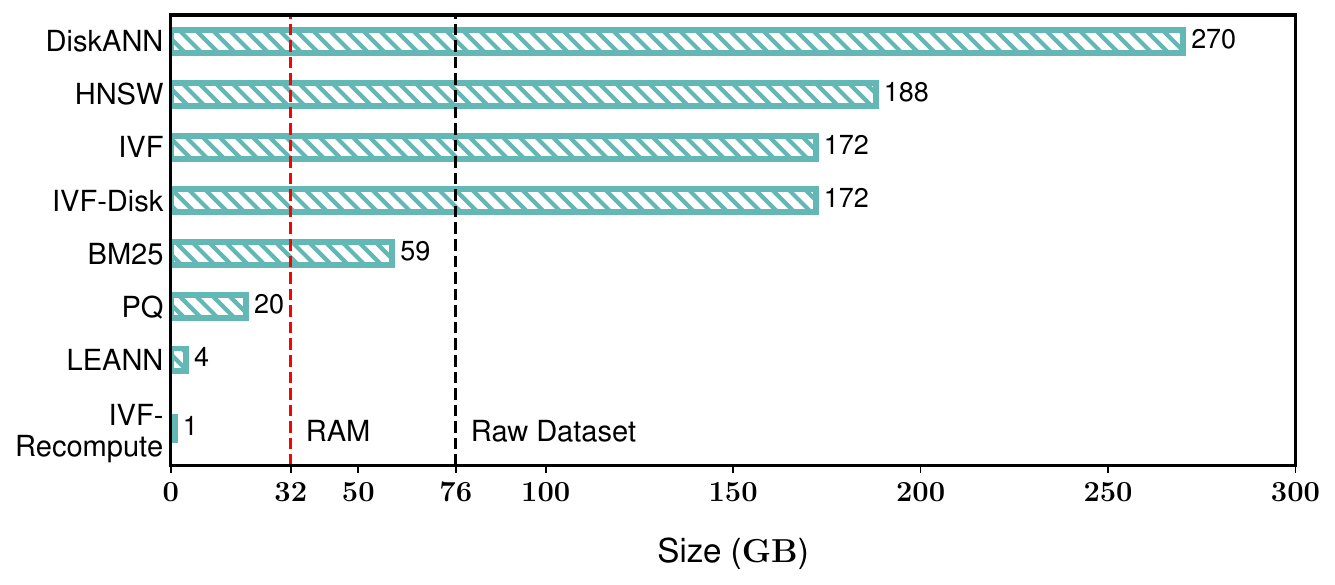}
	\caption{
		Storage consumption of different vector index methods on the RPJ-Wiki dataset. The black dashed line marks
 the raw dataset size (76 GB), while the red dashed line shows the typical RAM capacity (32 GB, RTX 4090, following our testbed configuration in \cref{sec:setup}). Memory-heavy methods like HNSW exceed this RAM limit and cannot run on such hardware. LEANN achieves the lowest storage footprint at only 5\% of the original dataset size.
	}
	\label{fig:storage-fig}
\end{figure}

We begin by describing the experimental setup in \cref{sec:setup}.
Then, in \cref{sec:main}, we present the main results and answer the following key questions:
(1) What is the storage overhead of different indexing methods?
(2) What is the latency of various vector search methods and the end-to-end RAG pipeline using them?
(3) What is the end-to-end RAG accuracy achieved by different methods?
Finally, in \cref{sec:ablation}, we conduct comprehensive ablation studies to evaluate the effectiveness of each component in \sys.

\subsection{Experiment Settings}
\label{sec:setup}

\begin{table*}[!ht]
	\centering
	\caption{Vector search and end-to-end RAG latency across datasets on RTX 4090. Latency is reported at 90\% recall. Results for PQ and BM25 are omitted as they fail to reach this accuracy (their downstream accuracy is also low in \autoref{fig:acc}). Results for HNSW and IVF are measured on a server with larger memory, as they cause OOM on the local RTX 4090. Overhead (\%) denotes the ratio of retrieval latency to total pipeline latency (retrieval / [retrieval + generation]).}
	\vspace{4pt}
	\renewcommand{\arraystretch}{1.25}
	\resizebox{\textwidth}{!}{
		\begin{tabular}{lcccc|lcccc}
			\toprule
			\textbf{Dataset}          & \textbf{Generation (s)} & \textbf{Method}                               & \textbf{Retrieval (s)}                & \textbf{Overhead (\%)}                 & \textbf{Dataset}       & \textbf{Generation (s)} & \textbf{Method}                               & \textbf{Retrieval (s)}                & \textbf{Overhead (\%)}                 \\
			\midrule

			\multirow{8}{*}{NQ}       & \multirow{8}{*}{20.86}
			                          & HNSW             & 0.05                                          & 0.20                                  & \multirow{8}{*}{GPQA}              & \multirow{8}{*}{69.60} & HNSW & 0.04                                          & 0.06                                  \\
			                          &                        & IVF                                     & 2.55                                  & 10.90                                  &                        &            & IVF                                     & 0.17                                  & 0.25                                   \\
			                          &                        & DiskANN                                       & 0.03                                  & 0.10                                   &                        &            & DiskANN                                       & 0.03                                  & 0.05                                   \\
			                          &                        & IVF-Disk                                      & 3.44                                  & 14.10                                  &                        &            & IVF-Disk                                      & 0.06                                  & 0.09                                   \\
			                          &                        & IVF-Recompute                                 & 307.61                                & 93.60                                  &                        &            & IVF-Recompute                                 & 21.88                                 & 23.20                                  \\
			                          &                        & PQ Compression                                & --                                    & --                                     &                        &            & PQ Compression                                & --                                    & --                                     \\
			                          &                        & BM25                                          & --                                    & --                                     &                        &            & BM25                                          & --                                    & --                                     \\
			                          &                        & \cellcolor[HTML]{EFEFEF}\textbf{LEANN} & \cellcolor[HTML]{EFEFEF}\textbf{2.48} & \cellcolor[HTML]{EFEFEF}\textbf{10.60} &                        &            & \cellcolor[HTML]{EFEFEF}\textbf{LEANN} & \cellcolor[HTML]{EFEFEF}\textbf{1.12} & \cellcolor[HTML]{EFEFEF}\textbf{1.60}  \\
			\midrule

			\multirow{8}{*}{TriviaQA} & \multirow{8}{*}{17.17}
			                          & HNSW             & 0.04                                          & 0.20                                  & \multirow{8}{*}{HotpotQA}          & \multirow{8}{*}{23.28} & HNSW & 0.05                                          & 0.20                                  \\
			                          &                        & IVF                                     & 3.54                                  & 17.10                                  &                        &            & IVF                                     & 3.87                                  & 14.20                                  \\
			                          &                        & DiskANN                                       & 0.06                                  & 0.30                                   &                        &            & DiskANN                                       & 0.11                                  & 0.50                                   \\
			                          &                        & IVF-Disk                                      & 3.65                                  & 17.50                                  &                        &            & IVF-Disk                                      & 5.05                                  & 17.80                                  \\
			                          &                        & IVF-Recompute                                 & 399.12                                & 95.90                                  &                        &            & IVF-Recompute                                 & 429.46                                & 94.80                                  \\
			                          &                        & PQ Compression                                & --                                    & --                                     &                        &            & PQ Compression                                & --                                    & --                                     \\
			                          &                        & BM25                                          & --                                    & --                                     &                        &            & BM25                                          & --                                    & --                                     \\
			                          &                        & \cellcolor[HTML]{EFEFEF}\textbf{LEANN} & \cellcolor[HTML]{EFEFEF}\textbf{2.96} & \cellcolor[HTML]{EFEFEF}\textbf{14.70} &                        &            & \cellcolor[HTML]{EFEFEF}\textbf{LEANN} & \cellcolor[HTML]{EFEFEF}\textbf{7.12} & \cellcolor[HTML]{EFEFEF}\textbf{23.40} \\
			\bottomrule
		\end{tabular}
	}
	\label{tab:rag_latency}
\end{table*}

\begin{table}[!ht]
	\centering
	\scriptsize
	\caption{Storage usage and retrieval latency overhead of \sys on personal datasets (RTX 4090).
		Overhead (\%) follows the definition in \cref{tab:rag_latency}, and Storage Savings (\%) denote the storage consumption of \sys relative to HNSW.}
	\vspace{2mm}
	\renewcommand{\arraystretch}{1.2}
	\begin{tabular}{@{}l@{\hspace{4pt}}c@{\hspace{4pt}}c@{\hspace{4pt}}c@{\hspace{4pt}}c@{}}
		\toprule
		\textbf{Dataset} & \textbf{Generation (s)} & \textbf{Retrieval (s)} & \textbf{Overhead (\%)} & \textbf{Storage Savings (\%)} \\
		\midrule
		FinanceBench     & 46.0                    & 1.5                    & 3                      & 97                            \\
		Enron            & 22.3                    & 1.9                    & 8                      & 98                            \\
		LAION            & 6.6                     & 1.6                    & 20                     & 97                            \\
		\bottomrule
	\end{tabular}
	\label{tab:rag-storage-saving-overhead-new-dataset}
\end{table}

\stitle{Workloads: Datastore and QA dataset.}
We construct the retrieval datastore using the RPJ-Wiki dataset~\cite{together2023redpajama}, a widely used corpus comprising approximately \textbf{76 GB} of raw Wikipedia text. Following prior work~\cite{shao2024scaling}, we segment the text into 256-token chunks and generate an embedding for each chunk using \textsc{Contriever} \cite{izacard2021unsupervised}, yielding 768-dimensional vectors. In total, we obtain 60 million ($N=60\text{M}$) passages, producing about 173 GB of embeddings. For the QA datasets, we adopt four standard benchmarks commonly used in RAG and open-domain retrieval: NQ~\cite{kwiatkowski-etal-2019-natural}, TriviaQA~\cite{joshi2017triviaqa}, GPQA~\cite{rein2024gpqa}, and HotpotQA~\cite{yang2018hotpotqa}.
Beyond the Wikipedia QA task, we further evaluate on FinanceBench~\cite{islam2023financebenchnewbenchmarkfinancial} for \textit{financial document retrieval}, the Enron Email Corpus~\cite{ryan2024enronqa} for \textit{email retrieval}, and LAION~\cite{schuhmann2021laion} for \textit{image data retrieval}.

\stitle{Testbed.}
We evaluate our system on two hardware platforms.
The first is a workstation with an NVIDIA RTX 4090 GPU~\cite{nvidia4090}, 32GB RAM, and a 1 TB disk running WSL2.
The second is an AWS EC2 M1 Mac instance~\cite{mac} with an Apple M1 Ultra (Arm64) processor, macOS, 128GB RAM and a 512 GB EBS volume.

\stitle{Baselines.}
We compare \sys against the following baselines:
\textbf{HNSW}~\cite{hnsw},
\textbf{IVF},
\textbf{DiskANN}~\cite{diskann},
\textbf{IVF-Disk},
\textbf{IVF-Recompute}~\cite{seemakhupt2024edgerag},
\textbf{PQ Compression}~\cite{pq}, and
\textbf{BM25}~\cite{msjimmy_bm25}.
These baselines cover graph-based, cluster-based, quantization-based, and lexical retrieval paradigms.
Detailed configurations are provided in \cref{app:baseline-details}.

\subsection{Main results}
\label{sec:main}
\stitle{Storage consumption.}
We compare the storage consumption of all baselines and \sys in \cref{fig:storage-fig}. Among all methods, only \sys and IVF-Recompute maintain total storage overhead below 5\% of the raw dataset size ($76$~GB).

Most existing systems incur substantial overhead, up to 2.5$\times$ the raw data size, making them impractical for deployment on \personaldevices. HNSW stores every dense embedding along with its graph connections, where each node contains a 768-dimensional embedding vector and padding for up to 60 neighbors (the maximum degree). DiskANN further amplifies this overhead due to its sector-aligned layout: each node's embedding ($768 \times 4$ bytes) and neighbor list ($60 \times 4$ bytes) are padded to 4KB SSD sectors. It also requires an additional 30GB for PQ embeddings, yielding the largest footprint ($270$~GB) among all methods. IVF and IVF-Disk exhibit similar overheads, both dominated by the cost of storing full embeddings. For BM25, the index size scales with the vocabulary and is roughly comparable to the raw corpus size in our setup. PQ compresses embeddings to a similar size as \sys ($5$GB) but requires an additional 15GB for graph index metadata. In contrast, \sys stores only a compact graph and highly compressed PQ embeddings, resulting in less than 5\% additional storage overhead. Among all baselines, IVF-Recompute achieves the smallest footprint by storing only IVF centroids on disk.

We also compare \sys against the most widely used HNSW index in \cref{tab:rag-storage-saving-overhead-new-dataset}, showing that it achieves over 97\% storage savings across diverse datasets.

\stitle{Latency evaluation for vector search and RAG.}
We evaluate the latency of vector search and end-to-end RAG across different methods in \cref{tab:rag_latency} and \cref{tab:rag-storage-saving-overhead-new-dataset}. We measure the per-request retrieval latency required to achieve 90\% recall (Recall@3, defined in \cref{sec:bg}) and the subsequent generation time. Detailed measurement procedures are provided in \crefrange{app:latency-measure}{app:latency-rag}.

First, we show that \textbf{\textit{second-level retrieval latency is acceptable}}. The generation phase dominates the total response time, typically exceeding 10s and reaching up to 70s, so the design choice in \sys to trade a small amount of latency for substantial storage savings is well justified.

Second, \textbf{\textit{while several methods achieve storage efficiency, only \sys delivers both high speed and accuracy.}} Among all baselines, only BM25, PQ, IVF-Recompute, and \sys maintain storage overhead below the size of the raw dataset, as shown in \cref{fig:storage-fig}. However, BM25 and PQ exhibit low retrieval accuracy and fail to reach 90\% recall. IVF-Recompute attains high recall but requires up to two orders of magnitude longer retrieval time than \sys (up to $200\times$ slower).

\begin{figure}[tbp]
	\centering
	\includegraphics[width=\columnwidth+0.25em]{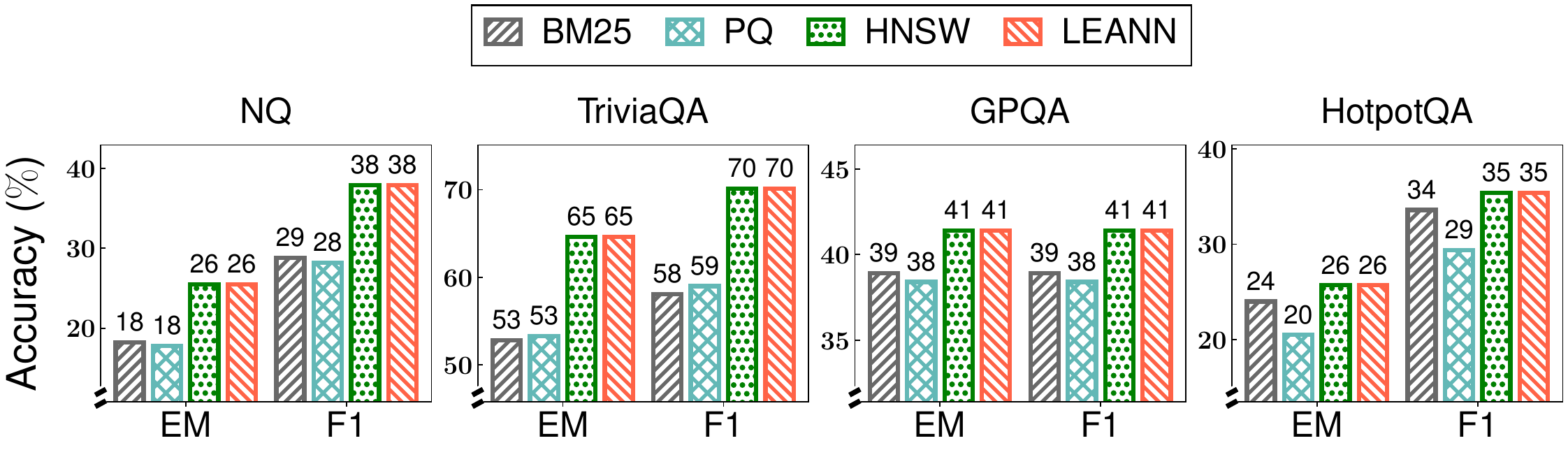}
	\caption{Comparison of Exact Match and F1 scores for downstream RAG tasks across four methods: keyword search (BM25), PQ-compressed vector search, HNSW, and LEANN. HNSW and \sys are configured to achieve a target recall of 90\%, while the PQ baseline is given extended search time to reach its highest possible recall. Here we use \textit{Qwen3-4B} as the generation model.
	}
	\label{fig:acc}
\end{figure}

This difference arises because \sys employs a graph-based index with $\mathcal{O}(\log N)$ embedding recomputation, while IVF-Recompute performs $\mathcal{O}(\sqrt{N})$ recomputations~\cite{wang2021comprehensive} ($N{=}60\text{M}$ in our experiments). Additional latency optimizations described in \cref{sec:latency_optimize} further contribute to \sys's performance advantage.

Finally, although \sys's standalone vector search latency is higher than graph-based baselines such as HNSW and DiskANN, we show that \textbf{\textit{\sys introduces negligible latency overhead when integrated into the full RAG pipeline}} on \personaldevices, while using far less storage. As shown in \cref{tab:rag_latency} and \cref{tab:rag-storage-saving-overhead-new-dataset}, \sys consistently adds less than 20\% latency overhead to the end-to-end retrieval and generation process. For reasoning-intensive tasks such as the graduate-level QA benchmark GPQA, the additional overhead introduced by \sys remains under 3\%, as the model’s long chain-of-thought generation dominates total latency.

We include Mac latency results in \autoref{tab:rag_latency_mac} (\cref{app:latency-mac}), using the same setup as \autoref{tab:rag_latency}.
Despite the Mac's lower TFLOPS, all previous conclusions hold, demonstrating \sys's generalization across platforms.

\stitle{Accuracy of downstream RAG applications.}
To evaluate RAG accuracy, we compare all retrieval methods on four QA datasets using Exact Match (EM) and F1 as metrics, as shown in \cref{fig:acc}. \sys achieves the highest downstream QA performance among all methods.

As shown in \cref{fig:acc}, \sys consistently outperforms BM25 and PQ across all datasets. It improves EM by up to 11.8\% over BM25 and 11.3\% over PQ, and F1 by up to 12.0\% and 11.1\%, respectively. The gains are most pronounced on factual answering benchmarks such as NQ and TriviaQA, where accurate semantic retrieval provides clear benefits. In contrast, the improvement is smaller on GPQA and HotpotQA. This is because RPJ-Wiki datastore is partially out-of-distribution for GPQA, which contains graduate-level questions that are less supported by Wikipedia content, and HotpotQA requires multi-hop reasoning, while our setup performs only single-hop retrieval.

Finally, when a target recall level (i.e., 90\%) is enforced, the downstream accuracy of \sys matches that of HNSW, confirming that our method preserves accuracy while achieving substantial storage savings.

\begin{figure}[!t]
	\centering
	\includegraphics[width=0.8\columnwidth]{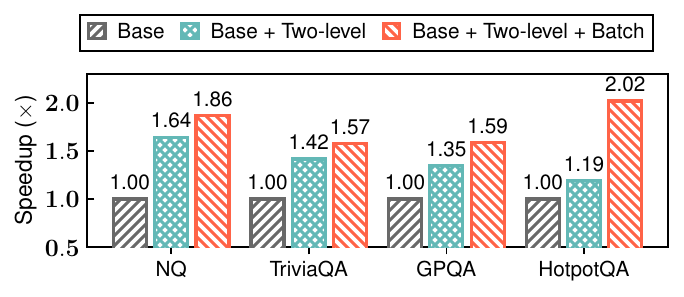}
	\caption{
		Speedup achieved by different optimization techniques described in \cref{sec:latency_optimize} when evaluated on four datasets to reach the same recall level. \textit{Two-level} refers to the optimization in \cref{sec:doubleq}, while \textit{Batch} corresponds to \cref{sec:batch}.
	}
	\label{fig:latecy_ablation}
\end{figure}

\subsection{Ablation Studies and Micro Benchmarks}
\label{sec:ablation}

\begin{figure}[!t]
	\centering
	\includegraphics[width=0.8\columnwidth]{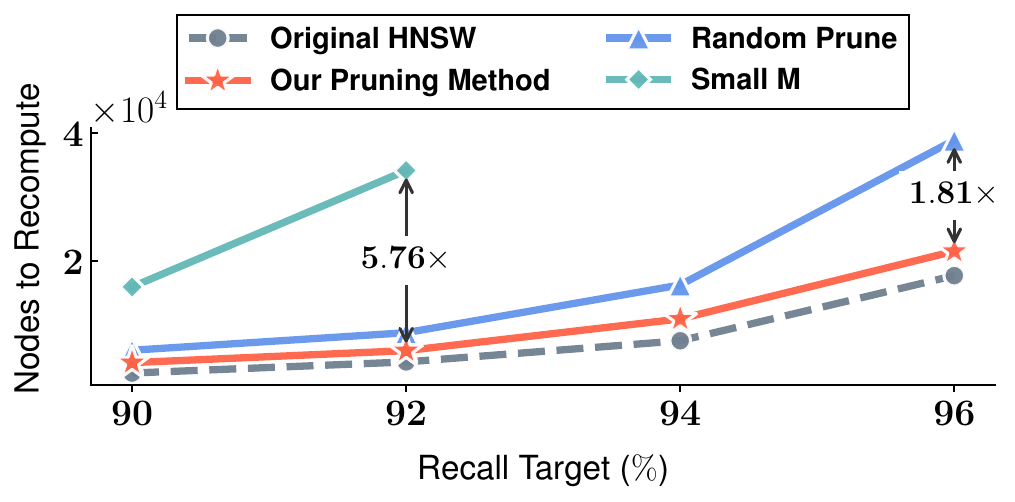}
	\caption{Comparison of pruned graph quality against two heuristic methods using the datastore in \cref{sec:setup}. At each recall target on the NQ dataset, we vary the search queue length \(ef\) to determine the minimum number of nodes that must be recomputed (lower is better). The dashed gray line represents the original HNSW graph as a baseline reference, which uses twice the storage (i.e., average degree) of the pruned methods.}
	\label{fig:graph_prune_ablation}
\end{figure}

\begin{figure}[!t]
	\centering
	\includegraphics[width=1\columnwidth]{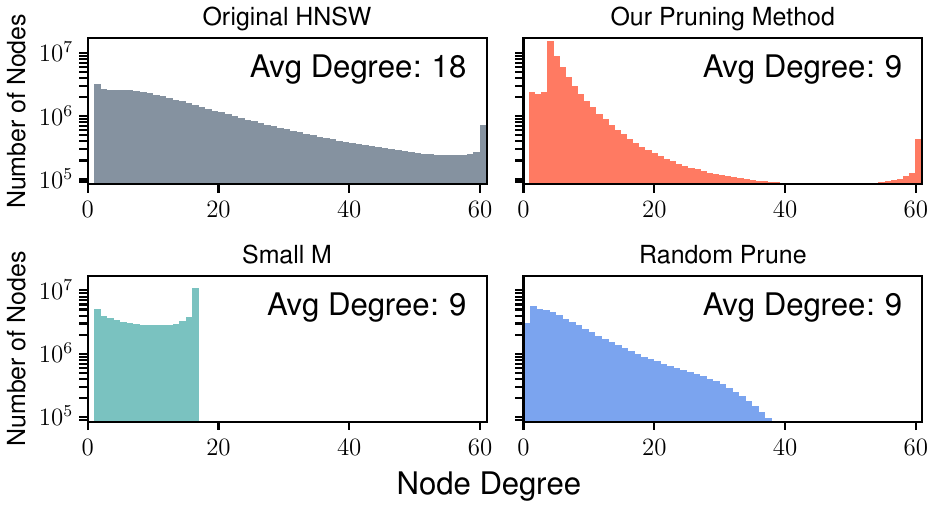}

	\caption{Comparison of (out-)degree distributions among the original graph, our pruning method, and two heuristic baselines. Similar to \cref{fig:graph_prune_ablation}, the gray curve represents the original HNSW graph, which has twice the size of the others. Only our pruning method successfully preserves the high-degree nodes.}
	\label{fig:degree_ablation}

	\Description{}
\end{figure}
\stitle{Latency optimization techniques.}  To evaluate the latency optimizations in \sys described in \cref{sec:latency_optimize}, we incrementally enable each component while maintaining a fixed target recall across multiple datasets.
Starting from a naive HNSW recomputation baseline, adding the \twolevel mechanism~(\cref{sec:doubleq}) yields an average $1.4\times$ speedup (up to $1.6\times$) by reducing the number of nodes requiring recomputation, with \sys enabling lightweight distance estimation without invoking the embedding generator.
Incorporating \dynbatch further improves GPU utilization during recomputation, increasing the average speedup to $1.8\times$ and the peak to $2.0\times$.
Among all datasets, HotpotQA gains the most from \dynbatch, as its longer search paths allow more effective grouping of multi-hop requests.

\stitle{Alternative \graphprune methods.}
We compare our \highdegreeprune algorithm with two baselines: \textit{(1) Random Prune}, which randomly removes 50\% of edges from the original graph; and \textit{(2) Small M}, which constrains the maximum out-degree during graph construction, yielding an average degree half that of the original graph. We evaluate graph quality by measuring the number of nodes that must be recomputed to achieve a given recall target, as shown in \cref{fig:graph_prune_ablation}. In \sys, latency is dominated by embedding recomputations, making this metric a proxy for retrieval latency.
The original graph has an average degree of 18. All three pruning methods, ours and the two baselines, are applied to reduce the average degree by half, from degree of 18 to 9, thereby halving the graph's storage overhead. As shown in \cref{fig:graph_prune_ablation}, our pruning method introduced in \cref{sec:pruning} achieves performance comparable to the original unpruned graph, while using only half the edges. To reach the same recall levels, Random Prune requires up to $1.8\times$ more nodes to recompute, while Small M requires up to $5.8\times$ more nodes to recompute. We omit the Small M results at the 94\% and 96\% recall targets, as it fails to reach these accuracy levels.

\stitle{Degree distribution in pruned graphs.}
To better understand the effectiveness of our pruning strategy, we analyze the out-degree distributions of the original graph, our approach, Random Prune, and Small M. As discussed in \cref{sec:pruning}, our design explicitly aims to preserve high-degree “hub” nodes. As shown in \cref{fig:degree_ablation}, it successfully retains a substantial number of such nodes, whereas the other two baselines fail to do so. This underscores the critical role of hub nodes in supporting efficient graph-based vector search, a finding that aligns with insights from prior work~\cite{hmann,munyampirwa2024down,skewmanohar2024parlayann}.

\begin{figure}[!t]
	\centering
	\includegraphics[width=\columnwidth]{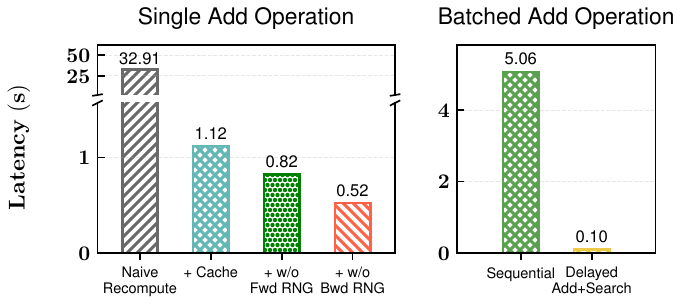}
	\caption{
		Comparison of update methods
	}
	\label{fig:storage_efficient_add}
\end{figure}

\stitle{Index update.} \autoref{fig:storage_efficient_add} shows how latency changes as we incrementally enable the optimizations of the \sys add operation described in \cref{sec:sys-add}.
We achieve up to a $63.3\times$ speedup over the naive method, consistent with our theoretical analysis.
On the right, introducing a buffer to delay batched additions further improves search speed while maintaining accuracy.

\stitle{More experiments.} We provide additional experiment results in \cref{app:more-ablation}.
In \cref{app:index-construct-compare}, we show that our \shardedmerge preserves the quality of proximity graph index while significantly reducing the peak storage during index-building .
In \cref{app:light-embed}, we show that using a smaller embedding model further accelerates \sys without compromising accuracy.
We also show that caching the exact embeddings for some hot objects effectively reduces query latency in \cref{app:disk-cache} and decompose the query latency of \sys in \cref{app:latency-breakdown}.

\section{Related Work}

\stitle{Resource-constrained vector search.}
Many works aim to reduce the memory cost of vector search. Disk-based systems like DiskANN~\cite{diskann} store vectors and graphs on disk with compressed in-memory embeddings for navigation. Starling~\cite{wang2024starling_sigmod} improves disk I/O, and FusionANNS~\cite{fast25} coordinates SSD, CPU, and GPU to lower cost. AiSAQ~\cite{tatsuno2024aisaq_gps_ref46} and LM-DiskANN~\cite{LM-DiskANN} further cut DRAM use by keeping compressed embeddings on disk.
EdgeRAG~\cite{seemakhupt2024edgerag} generates embeddings online via an IVF-based index but still suffers high storage and recomputation overhead. 
MicroNN~\cite{zhang2025micronn} and ObjectBox~\cite{objectbox_docs} are optimized for personal devices but still require storing all embeddings.
Embedding compression methods like PQ~\cite{pq} and RabitQ~\cite{gao2024rabitq_gps_ref15} save space but lose accuracy under tight budgets. In contrast, \sys combines on-the-fly embedding recomputation with a pruned graph index and optimized traversal for \personaldevices.

\stitle{Vector search applications on \personaldevices.} On-device vector search enables privacy-preserving, low-latency, and offline capabilities across diverse applications. On-device RAG systems ground language models in personal document collections while maintaining data privacy~\cite{ryan2024enronqa, wang2024mememo, work-in-progress, zerhoudi2024personarag}. Personalized recommenders~\cite{yin2024devicers} match user profiles with item embeddings on the device, while vision-based search~\cite{snap_cvpr2023_tutorial} retrieves local images or videos to assist downstream QA or generation tasks. These applications motivate the design of \sys to enable efficient, low-overhead vector search on \personaldevices.

\section{Conclusions}
\label{sec:conclusion}
Similarity search over high-dimensional embeddings underpins many generative AI applications such as RAG. 
However, enabling such capabilities remains challenging due to the substantial storage required for embeddings and index metadata. 
We present \sys, a storage-efficient vector index based on \textit{graph-based recomputation}. 
By combining \textit{two-level search} with \textit{dynamic batching}, \sys supports efficient query processing without storing the full embedding set. 
A \textit{high-degree preserving pruning} strategy further reduces graph storage while maintaining accuracy. 
\sys also offers fast, storage-efficient index construction and update pipelines. 
Together, these techniques allow \sys to operate with an index smaller than 5\% of the raw data size, achieving up to 50$\times$ storage reduction compared to existing methods while preserving high recall and low latency.

\bibliography{example_paper}
\bibliographystyle{mlsys2025}

\appendix
\clearpage
\section{RNG Pruning}
\label[appendix]{sec:RNG}

\begin{figure}[htbp]
	\centering
	\includegraphics[width=\columnwidth]{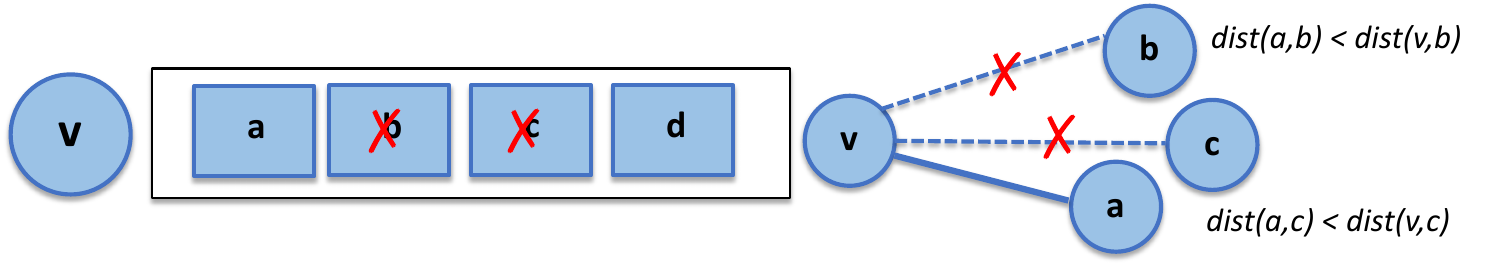}
	\vspace{-1mm}
	\caption{Select neighbors from candidate nodes using RNG.
}
	\label{fig:rng}
	\vspace{-1mm}
	\Description{}
\end{figure}

For a node \(v\) inserted into any proximity-graph index (including HNSW), the algorithm first searches for a list of candidate neighbors \(a, b, c, d\) (see \cref{alg:best-search}) sorted by distance from \(v\). 
The RNG-based pruning rule\cite{rng,Toussaint1980TheRN}, implemented in Line~\ref{line:RNG},  then iterates through this list in order of increasing distance. 
A candidate \(x\) is pruned if there exists a closer neighbor \(a\) such that \(\text{Dist}(a, x) < \text{Dist}(v, x)\). 
This effectively removes the longest edge in the triangle formed by \((v, a, x)\). 
As illustrated in \cref{fig:rng}, the edges \(v\!-\!b\) and \(v\!-\!c\) are pruned, so the search from \(v\) to \(b\) or \(c\) proceeds indirectly through \(a\).
This pruning strategy is widely used in modern graph-based ANN indexes and makes the resulting graph extremely sparse~\cite{munyampirwa2024down,hnsw}.

\section{\sys Update Strategy}
\label[appendix]{sec:sys-add}

The \textsc{Add} algorithm is presented in \cref{alg:add}.
\subsection{Add Operation: Method and Time Complexity}
\stitle{Naive Implementation.}
A naive implementation of the \textsc{Add} operation in \sys recomputes all distances from scratch since only the graph structure is stored. 
Its total time complexity can be expressed as:
\[
O(M \cdot efC + efC^2 + M^3),
\]
where \(efC\) is the construction queue length and \(M\) the maximum node degree.

We first analyze the complexity of \textsc{ShrinkNeighborList}. 
Each node placed in the retained set \(R\) may be re-examined up to \(|W|\) times, giving a cost of \(O(|W|^2)\) for a single call.

Specifically:
\begin{itemize}
    \item \textsc{SearchNeighborsToAdd} performs a one-time neighbor search without revisiting nodes, yielding \(O(M \cdot efC)\). 
    Caching offers no benefit since nodes are not revisited.
    \item \textsc{ShrinkNeighborList} (Line~\ref{line:shrink-neighbor}) runs in \(O(efC^2)\), as it computes pairwise distances among up to \(efC\) candidates.
    \item Adding forward edges requires no recomputation, since each node maintains at most \(M\) links.
    \item Adding reverse edges costs \(O(M^3)\), as up to \(M\) neighbors are updated and each triggers an \(O(M^2)\) RNG-based shrink.
\end{itemize}

\stitle{Caching Optimization.}
To improve efficiency, \sys introduces a distance cache to eliminate redundant computations in the \textsc{Shrink} step, reducing the overall complexity to:
\[
O(M \cdot efC + efC + M^2),
\]
since the shrink operation now costs only \(O(|W|)\).

\stitle{Simplified RNG Pruning.}
By further simplifying \textsc{ShrinkNeighborList} to randomly select neighbors instead of performing full RNG checks, the complexity becomes:
\[
O(M \cdot efC + M^2).
\]
Finally, applying the same simplification to the reverse-edge update step yields the optimized complexity:
\[
O(M \cdot efC),
\]
reducing the cost from cubic to linear in \(M\) while maintaining comparable graph connectivity.

\subsection{Batched Add Operation: Optimization}
When a batch of add operations is followed by a search request, \sys does not immediately insert all new passages. 
Instead, it temporarily buffers their embeddings and merges search results from both the existing graph and the buffered embeddings. 
After the search completes, the buffered passages are inserted \textit{asynchronously}, a process we term \textit{delayed insertion}.

The same system optimizations described earlier can be reused here, with the addition of a global cache to avoid redundant computations across multiple add requests. 
To maintain storage efficiency, \sys monitors the cache size and clears it once a predefined budget is reached, starting a new round of batched insertion.

\subsection{Soft Deletion Strategy}
\label[appendix]{sec:soft-delete}
\sys adopts a simple soft delete for graph nodes. Each node keeps a binary delete flag, so removal is an $O(1)$ update that leaves the adjacency list untouched. During query processing, we still traverse deleted nodes to reach their neighbors, but before producing results we filter the candidate queue (the $EQ$ in \cref{alg:two_level}) by this flag and then take the top-$k$ active entries to guarantee correctness.

If the fraction of deleted nodes grows beyond a threshold (e.g., 5\%), we can trigger a background rebuild. Exploring such policies is left for future work.

\begin{algorithm}[t]
    \caption{Node Insertion into Graph Index}
    \begin{algorithmic}[1]
        \State \textbf{Input:} Existing graph \(G\); construction queue length \(efC\); maximum degree \(M\); node to insert \(v\)
        \State \textbf{Output:} Updated graph \(G\) including node \(v\)
\Function{Shrink}{$W, M$}
    \State Initialize $R \gets \emptyset$
    \For{$x$ in $W$ (by ascending distance to query $q$)}
        \If{no $y \in R$ s.t. $\text{Dist}(x, y) < \text{Dist}(x, q)$}\label{line:RNG}
            \State Add $x$ to $R$
        \EndIf
        \If{$|R| = M$}
            \State \textbf{break}
        \EndIf
    \EndFor
    \State \Return $R$
\EndFunction
        \State \(W \gets \textsc{Search}(v, efC)\) \Comment{See \cref{alg:best-search}}
        \State \(W \gets \textsc{Shrink}(W, M)\) \label{line:shrink-neighbor} \Comment{ShrinkNeighborList}
        \State Add directed edges from \(v\) to all nodes in \(W\)
\For{each \(w \in W\)}
    \State Add directed edge from \(w\) to \(v\)
    \If{\(\deg(w) > M\)}
        \State \(\textsc{Shrink}(w\text{'s neighbor list}, M)\)
    \EndIf
\EndFor
    \end{algorithmic}
    \label{alg:add}
\end{algorithm}

\section{Evaluation details}
\subsection{Baseline Configurations}
\label[appendix]{app:baseline-details}

\begin{itemize}[leftmargin=*]

    \item \textbf{HNSW}~\cite{hnsw}: We use the \texttt{faiss.IndexHNSWFlat} implementation with construction parameters recommended by FAISS: $M{=}30$ and $efConstruction{=}128$, distinct from the search-time parameter $ef$. 

    \item \textbf{IVF}~\cite{ivf}: We adopt the \texttt{faiss.IndexIVFFlat} implementation. Following best practices from FAISS~\cite{faissGuidelines} and prior work~\cite{ivf_crtpt:2023/1438}, we set the number of centroids to $\sqrt{N}$, where $N$ is the size of the datastore. For our $N{=}60\text{M}$ setup, this corresponds to $nlist{=}8192$.

\item \textbf{DiskANN}~\cite{diskann}: We use DiskANN with $M{=}60$ and $efConstruction{=}128$, following recommended settings~\cite{diskann}. It stores only a PQ table in memory and loads full embeddings from disk on demand.

    \item \textbf{IVF-Disk}: Reduces memory usage by employing memory-mapped files (\texttt{mmap}) instead of loading the entire index into memory. Implemented using FAISS's \texttt{faiss.contrib.ondisk} module with the same parameters as IVF.

\item \textbf{IVF-Recompute}~\cite{seemakhupt2024edgerag}: Inspired by Edge-RAG, this variant recomputes embeddings at query time instead of storing them, using the same construction parameters as IVF.

\item \textbf{PQ Compression}~\cite{pq}: Applies Product Quantization to compress stored embeddings while preserving the graph structure. For fair comparison, we compress the vectors to 5~GB—slightly larger than our system's 4~GB footprint. We use the PQ implementation from~\cite{diskann}.

\item \textbf{BM25}~\cite{msjimmy_bm25,rekabsaz2021tripclick}: A classical lexical ranking method widely used in keyword-based retrieval systems. We employ the standard implementation from Pyserini~\cite{pyserini}.

\end{itemize}
\subsection{Latency Measurement and Evaluation Protocol}
\label[appendix]{app:latency-measure}

To evaluate retrieval accuracy, we report Recall@k as defined in \cref{sec:bg}.
In open-domain settings, ground-truth labels for retrieved passages are typically unavailable.  
Following standard practice~\cite{pq,schuhmann2021laion,zhu}, we treat the results from exact search as a proxy for ground truth.  
In all experiments, we set $k{=}3$, consistent with prior work~\cite{shao2024scaling,asai2023self}, and report Recall@3 as our retrival accuracy metric.

For latency evaluation, we measure the time required to achieve different target recall levels.  
Specifically, we perform a binary search to find the minimal search queue length $ef$ (as defined in \cref{alg:best-search}) that reaches the desired recall.
Using the resulting $ef$, we record the average retrieval latency over 20 random queries.  

\subsection{Latency Measurement in RAG Pipeline}
\label[appendix]{app:latency-rag}
We evaluate the latency of \sys at the 90\% recall level across all datasets. For text generation, we use \textit{Qwen3-4B}~\cite{yang2025qwen3}, and for multimodal workloads, we use \textit{Qwen2.5-VL-7B-Instruct}~\cite{bai2025qwen2}. Both the embedding and generation models are implemented using the Hugging Face framework.

\subsection{RAG Latency on Mac Platform}
\label[appendix]{app:latency-mac}
To validate the generalizability of our results across different hardware platforms, we conducted additional experiments on Mac hardware. \Cref{tab:rag_latency_mac} presents the vector search and end-to-end RAG latency measurements on Mac, following the same experimental protocol as the RTX 4090 results shown in the main paper. The results demonstrate that \sys maintains its efficiency advantages on Mac hardware, with retrieval overhead remaining low compared to other methods despite the different underlying architecture and computational characteristics.

\begin{table*}[!ht]
	\centering
\caption{Vector search and end-to-end RAG latency across datasets on Mac. Latency is reported at 90\% recall. Results for PQ and BM25 are omitted as they fail to reach this accuracy (their downstream accuracy is also low in \cref{fig:acc}). Results for HNSW and IVF are omitted, as they cause OOM on the local Mac and on all EC2 Mac instances on AWS. Overhead (\%) denotes the ratio of retrieval latency to total pipeline latency (retrieval / [retrieval + generation]).}
	\vspace{4pt}
	\renewcommand{\arraystretch}{1.25}
	\resizebox{\textwidth}{!}{
		\begin{tabular}{lcccc|lcccc}
			\toprule
			\textbf{Dataset}          & \textbf{Generation (s)} & \textbf{Method}                               & \textbf{Retrieval (s)}                & \textbf{Overhead (\%)}                 & \textbf{Dataset}       & \textbf{Generation (s)} & \textbf{Method}                               & \textbf{Retrieval (s)}                & \textbf{Overhead (\%)}                 \\
			\midrule

			\multirow{8}{*}{NQ}       & \multirow{8}{*}{45.42}
			                          & HNSW                   & --                                            & --                                    & \multirow{8}{*}{GPQA}              & \multirow{8}{*}{132.24} & HNSW       & --                                            & --                                    \\
			                          &                        & IVF                                           & --                                    & --                                     &                        &            & IVF                                           & --                                    & --                                     \\
			                          &                        & DiskANN                                       & 0.37                                  & 0.8                                    &                        &            & DiskANN                                       & 0.29                                  & 0.2                                    \\
			                          &                        & IVF-Disk                                      & 2.94                                  & 6.1                                    &                        &            & IVF-Disk                                      & 0.11                                  & 0.1                                    \\
			                          &                        & IVF-Recompute                                 & 2446.60                               & 98.2                                   &                        &            & IVF-Recompute                                 & 174.06                                & 56.8                                   \\
			                          &                        & PQ Compression                                & --                                    & --                                     &                        &            & PQ Compression                                & --                                    & --                                     \\
			                          &                        & BM25                                          & --                                    & --                                     &                        &            & BM25                                          & --                                    & --                                     \\
			                          &                        & \cellcolor[HTML]{EFEFEF}\textbf{LEANN} & \cellcolor[HTML]{EFEFEF}\textbf{13.84} & \cellcolor[HTML]{EFEFEF}\textbf{23.4} &                        &            & \cellcolor[HTML]{EFEFEF}\textbf{LEANN} & \cellcolor[HTML]{EFEFEF}\textbf{5.22} & \cellcolor[HTML]{EFEFEF}\textbf{3.8} \\
			\midrule

			\multirow{8}{*}{TriviaQA} & \multirow{8}{*}{52.92}
			                          & HNSW                   & --                                            & --                                    & \multirow{8}{*}{HotpotQA}          & \multirow{8}{*}{44.67} & HNSW       & --                                            & --                                    \\
			                          &                        & IVF                                           & --                                    & --                                     &                        &            & IVF                                           & --                                    & --                                     \\
                          &                        & DiskANN                                       & 0.98                                  & 1.8                                    &                        &            & DiskANN                                       & 1.91                                  & 4.1                                    \\
                          &                        & IVF-Disk                                      & 2.64                                  & 4.8                                    &                        &            & IVF-Disk                                      & 3.54                                  & 7.4                                    \\
                          &                        & IVF-Recompute                                 & 3174.41                               & 98.4                                   &                        &            & IVF-Recompute                                 & 3415.74                               & 98.7                                   \\
			                          &                        & PQ Compression                                & --                                    & --                                     &                        &            & PQ Compression                                & --                                    & --                                     \\
			                          &                        & BM25                                          & --                                    & --                                     &                        &            & BM25                                          & --                                    & --                                     \\
                          &                        & \cellcolor[HTML]{EFEFEF}\textbf{LEANN} & \cellcolor[HTML]{EFEFEF}\textbf{17.15} & \cellcolor[HTML]{EFEFEF}\textbf{24.5} &                        &            & \cellcolor[HTML]{EFEFEF}\textbf{LEANN} & \cellcolor[HTML]{EFEFEF}\textbf{44.80} & \cellcolor[HTML]{EFEFEF}\textbf{50.1} \\
			\bottomrule
		\end{tabular}
	}
	\label{tab:rag_latency_mac}
\end{table*}

\section{More Ablation Studies}\label[appendix]{app:more-ablation}

\subsection{Comparison of Index Construction}
\begin{figure}[!t]
	\centering
	\includegraphics[width=\columnwidth]{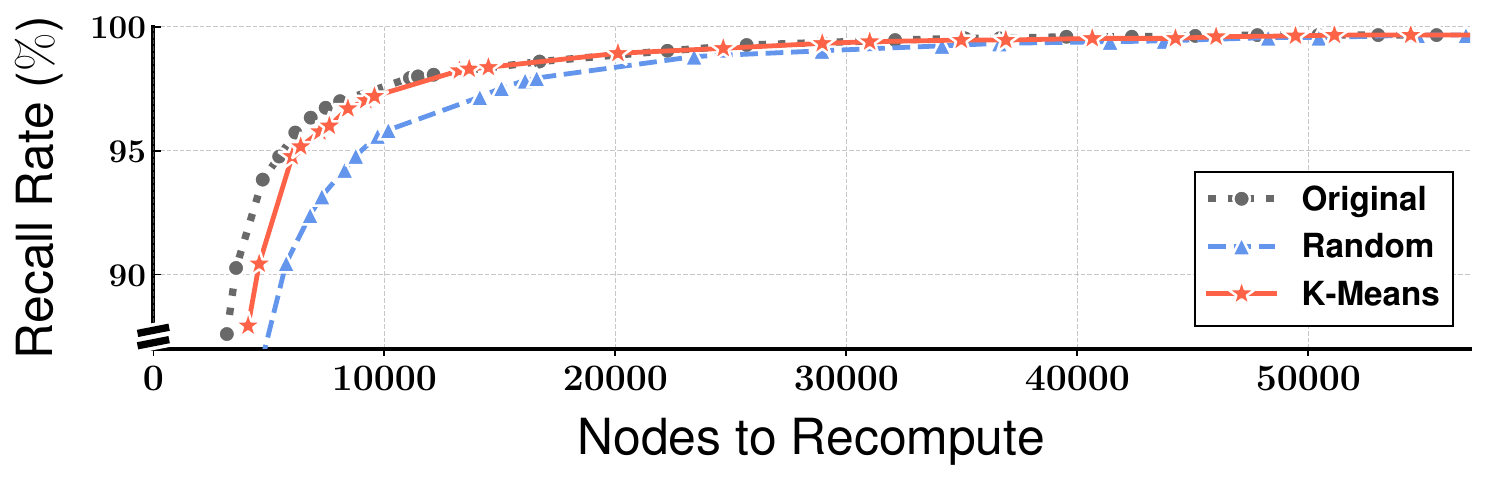}
	\caption{\textbf{[Ablation Study]:} Comparison of \storageeff index construction methods with the original HNSW.}
	\label{fig:effi-build}
\end{figure}

\label[appendix]{app:index-construct-compare}
We evaluate the \storageeff index construction techniques introduced in \cref{sec:build}, comparing them against the standard HNSW construction. 
Specifically, we implement two variants of the proposed \shardedmerge approach: 
(1) a \textit{k-means}–based method (see \cref{sec:build}) that groups similar passages before sharding, and 
(2) a random assignment baseline that omits clustering. 
We assess graph quality using the same methodology as before, with results shown in \cref{fig:effi-build}.
In this setup, the dataset is partitioned into 15 shards, achieving about a \textbf{5×} reduction in peak storage usage during index construction.

The \textit{k-means}–sharded graph achieves nearly the same recall as the original HNSW with only a small increase in recomputation cost, indicating that it maintains strong connectivity after sharding and merging. 
In contrast, the randomly sharded graph requires much more recomputation to reach the same recall. 
These results highlight the benefit of clustering similar passages before sharding, validating the design of our \shardedmerge approach.

\subsection{Using Different Embedding Model Sizes}
\begin{figure}[tbp]
    \centering
    \begin{subfigure}[t]{0.48\columnwidth}
        \centering
        \includegraphics[width=\linewidth]{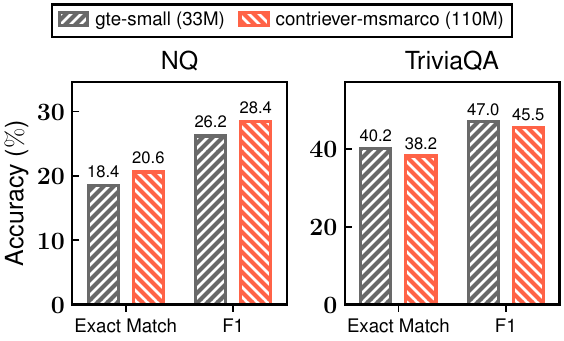}
        \caption{Accuracy\label{fig:small_accuracy}}
    \end{subfigure}
    \hfill
    \begin{subfigure}[t]{0.48\columnwidth}
        \centering
        \includegraphics[width=\linewidth]{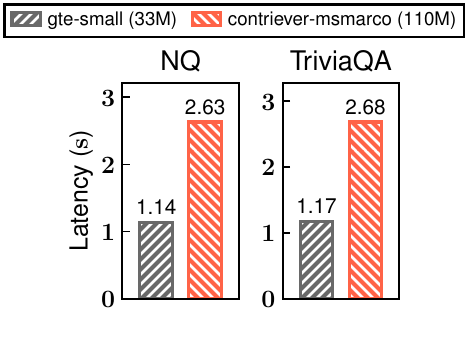}
        \caption{Latency\label{fig:small_latency}}
    \end{subfigure}
    \caption{\textbf{[Ablation Study]:} Downstream accuracy and end-to-end latency when swapping the embedding model for a lightweight alternative on a 2M-chunk datastore with \texttt{ef=50}.}
    \label{fig:small_embedd}
    \Description{Two subplots comparing downstream accuracy and latency for Contriever and GTE-small embeddings.}
\end{figure}

\label[appendix]{app:light-embed}
Since the primary bottleneck of our system lies in the recomputation process, we further explore the potential for latency reduction by adopting a smaller embedding model. Specifically, we replace the original \textit{Contriever} model (110M parameters) used in \cref{sec:setup} with the lightweight \textit{GTE-small} model~\cite{li2023towardsgte}, which has only 34M parameters. We evaluate performance on a 2M document datastore using a fixed search queue length of \texttt{ef=50}. As shown in \cref{fig:small_embedd}, \textit{GTE-small} achieves a $2.3\times$ speedup while maintaining downstream task accuracy within 2\% of the Contriever baseline, demonstrating that \sys can further reduce latency by leveraging lighter encoders without sacrificing answer quality.

\subsection{Relaxing Disk Constraint}
\begin{figure}[tbp]
    \centering
    \includegraphics[width=\columnwidth]{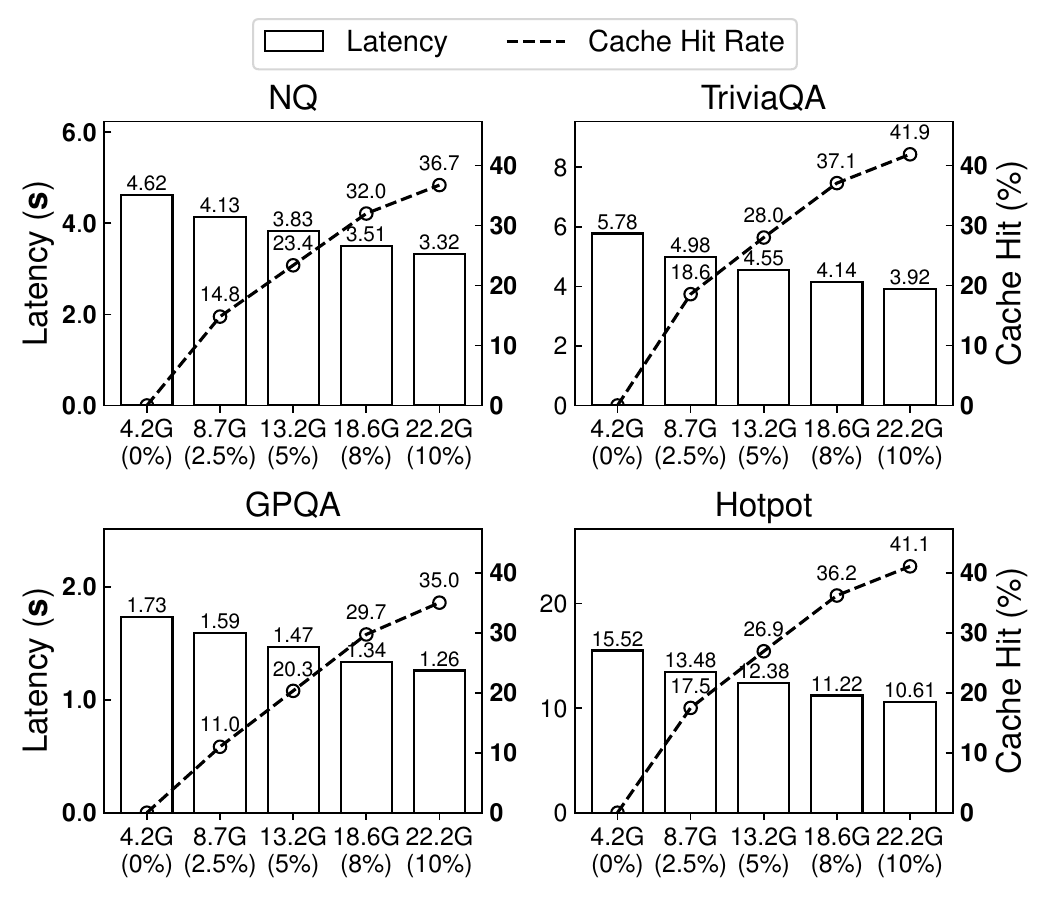}
    \caption{\textbf{[Ablation Study]:} Latency and cache-hit rate comparison under varying storage budgets.}
    \label{fig:disk_cache}
    \Description{Line chart showing how latency and cache-hit rate change as more embeddings are cached on disk.}
\end{figure}

\label[appendix]{app:disk-cache}
When disk storage constraints are relaxed, \sys can materialize the embeddings of high-degree nodes to reduce recomputation overhead. \Cref{fig:disk_cache} quantifies the resulting latency improvements and cache-hit rates across four datasets while varying the fraction of cached embeddings. Storing just 10\% of the original embeddings yields a $1.5\times$ speedup, with a cache hit rate of up to 41.9\%. This high cache hit rate arises from the skewed access pattern characteristic of graph-based traversal, though SSD loading overhead prevents the latency gains from matching the hit rate exactly.

\subsection{Graph-based Recomputation Breakdown}
\begin{figure}[tbp]
    \centering
    \includegraphics[width=\columnwidth]{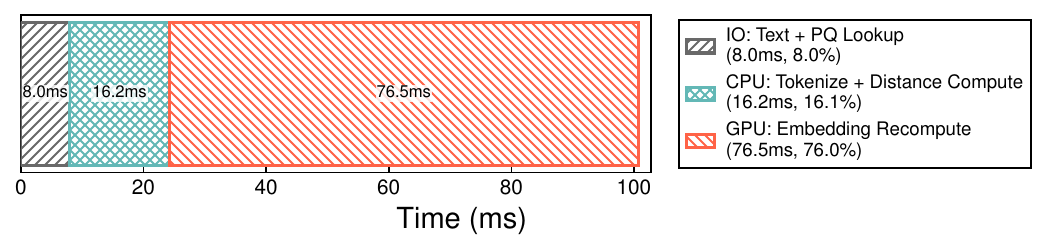}
    \caption{\textbf{[Ablation Study]:} Latency breakdown of a batch of requests during graph-based recomputation.}
    \label{fig:decouple}
    \Description{Stacked bar chart showing the proportion of latency spent on PQ lookup, text processing, and embedding recomputation.}
\end{figure}

\label[appendix]{app:latency-breakdown}
\Cref{fig:decouple} decomposes the latency of a batched query into three stages: PQ lookup, text processing, and embedding recomputation. Each batch aggregates multiple hops of recomputation, as described in \cref{sec:batch}. First, \sys performs PQ lookups to select promising nodes, then retrieves and tokenizes the corresponding raw text. The tokenized inputs are sent to the local embedding generator. Finally, \sys performs embedding recomputation and distance calculation. Although embedding recomputation is the primary bottleneck in \sys, accounting for roughly 76\% of total latency, the three stages span I/O, CPU, and GPU resources, indicating opportunities to overlap work and further improve efficiency.

\end{document}